\documentclass[twocolumn,byrevtex,prl,showpacs]{revtex4}

\usepackage{graphicx}
\usepackage{dcolumn}
 
\begin{document}

\title{Spectral function of the one-dimensional Hubbard model away
from half filling}

\author{H. Benthien}
\author{F. Gebhard}
\affiliation{Fachbereich Physik, Philipps-Universit\"{a}t, 
D-35032 Marburg, Germany}

\author{E. Jeckelmann}
\affiliation{Institut f\"{u}r Physik, Johannes Gutenberg-Universit\"{a}t, 
D-55099 Mainz, Germany}

\date{\today}

\begin{abstract}
We calculate the photoemission spectral function of the one-dimensional 
Hubbard model away from half filling using the dynamical density matrix
renormalization group method.
An approach for calculating momentum-dependent quantities
in finite open chains is presented.
Comparison with exact Bethe Ansatz results demonstrates the
unprecedented accuracy of our method.
Our results show that the
photoemission spectrum of the quasi-one-dimensional conductor
TTF-TCNQ provides evidence for spin-charge separation
on the scale of the conduction band width.
\end{abstract}


\maketitle

The Luttinger liquid theory describes the ground state
and asymptotic low energy properties of
one-dimensional correlated metals~\cite{Schoenhammer}.
Two characteristics of a Luttinger liquid are the absence
of quasi-particles predicted by the Fermi liquid theory
of normal metals and the occurrence of independent
spin and charge excitations.
In principle, these features can be observed
in the spectral function~\cite{Voit,Meden} which corresponds
to the spectrum measured in 
angle-resolved photoemission spectroscopy (ARPES)
experiments. 
In real materials, however, the low-energy properties
are likely to be governed by three-dimensional physics.
One-dimensional physics is observed only above
a crossover energy scale, even in the most strongly
anisotropic materials.
Consequently, it has proven difficult to observe 
unambiguous evidence for Luttinger liquid physics 
in experiments probing only low-energy properties of 
quasi-one-dimensional conductors. 

A recent ARPES experiment
for the quasi-one-dimen\-sional organic conductor TTF-TCNQ
(tetrathiafulvalene tetracyanoquinodimethane)
has revealed significant discrepancies from 
the predictions of Fermi liquid theory and
conventional electronic structure calculations~\cite{Claessen,Sing}.
The experimental spectrum dispersion can be consistently
mapped over the scale of the conduction band width
onto separated spin and charge excitation bands of
the one-dimensional Hubbard model~\cite{Hubbard} away from
half filling.
This is one of the strongest pieces of experimental evidence for
spin-charge separation and thus for 
Luttinger liquid physics in low-dimensional materials.  
However, a direct comparison of the experimental
ARPES spectrum with the Hubbard model spectral function
has not been possible yet.

The Hubbard model was solved exactly 36 years ago~\cite{Lieb}
and the dispersion of its excitation 
bands can be computed~\cite{Schulz}.
Nevertheless, the photoemission spectral function
can only be calculated exactly
in the limiting cases of noninteracting
electrons or infinitely strong electron 
interaction~\cite{Penc,Favand} and 
in the low-energy limit described by the
Luttinger liquid theory. 
Various numerical methods have provided a qualitative picture of 
spectral functions in the Hubbard model but
exact diagonalizations~\cite{Favand,Bannister}
are limited to too small systems
to investigate the thermodynamic limit
while other approaches~\cite{QMC,Senechal}
are based on various approximations of uncertain accuracy. 

We have determined the photoemission spectral function of the
one-dimensional Hubbard model with parameters appropriate
for TTF-TCNQ using the dynamical
density-matrix renormalization group (DDMRG) method~\cite{ddmrg}.
A novel approach is used to calculate momentum-dependent
quantities in finite open chains.
This allows us to investigate large systems almost exactly
and to make a direct comparison of the Hubbard model
spectral weight distribution with the experimental
TTF-TCNQ spectrum.
To demonstrate the accuracy of our method and to identify 
excitations contributing to the photoemission spectral function we
compare our numerical results with exact Bethe Ansatz
results.

A minimal model to describe the electronic properties
of TTF-TCNQ is the one-dimensional Hubbard model defined by the Hamiltonian
\begin{eqnarray}
\hat{H} 
&=& -t \sum_{l;\sigma} \left( \hat{c}_{l,\sigma}^+\hat{c}_{l+1,\sigma} 
+ \hat{c}_{l+1,\sigma}^+\hat{c}_{l,\sigma} \right) \nonumber \\
%
&&+ U \sum_{l} \hat{n}_{l,\uparrow}
\hat{n}_{l,\downarrow} - \mu \sum_{l} \hat{n}_l
\label{Hamiltonian}
\end{eqnarray}
Here $\hat{c}^+_{l,\sigma}$, $\hat{c}_{l,\sigma}$ are creation and 
annihilation operators for electrons with spin 
$\sigma = \uparrow,\downarrow$ at site $l=1,\dots,L$ (representing
a $\pi$-type Wannier orbital centered on a TCNQ molecule), 
$\hat{n}_{l,\sigma}= \hat{c}^+_{l,\sigma}\hat{c}_{l,\sigma}$, 
and $\hat{n}_l=\hat{n}_{l,\uparrow}+\hat{n}_{l,\downarrow}$.
Appropriate parameters for TTF-TCNQ  are an on-site Coulomb repulsion
$U=4.9t$ and  a hopping integral $t=0.4eV$~\cite{Claessen,Sing}.
(These values are appropriate for the TTF-TCNQ surface,
which is probed in ARPES experiments, not for bulk TTF-TCNQ.)
Although the filling of the TCNQ band is $n=0.59$,
we use a slightly different filling
$n=0.6$ in our simulations 
to facilitate the finite-size-scaling analysis. 
For a chain with $L$ sites and $N=nL$ electrons
the chemical potential $\mu$ is chosen so that 
$E_0(N-1) = E_0(N+1)$, where $E_0(N\pm1)$ is
the ground state energy with $N\pm1$ electrons.
Thus the Fermi energy is $\epsilon_F=0$ in the thermodynamic limit
$L\rightarrow \infty$.

The photoemission spectral function $A(k,\omega)$
is the imaginary part of the
one-particle Green's function
\begin{equation}
A(k,\omega) = \frac{1}{\pi} \Im
\langle \psi_0| \hat{c}_{k,\sigma}^{\dag} 
\frac{1}{\hat{H}+\omega-E_0-i\eta} \hat{c}_{k,\sigma}
| \psi_0 \rangle  ,
\end{equation}
where $|\psi_0\rangle$ and $E_0$ are the ground state wavefunction and energy 
of the Hamiltonian~(\ref{Hamiltonian}).
This function can be calculated for finite broadening $\eta$
and system sizes $L$ 
using the dynamical DMRG method~\cite{ddmrg}.
The spectral properties in the thermodynamic limit
can be determined using a finite-size-scaling 
analysis~\cite{ddmrg} with an appropriate broadening $\eta(L)$. 
Here we have used $\eta L = 9t$ and system
sizes up to $L=150$ sites.
DMRG truncation errors are negligible
for all results presented here
(up to $m= 400$ density-matrix eigenstates
have been kept per block in our calculations.)

The operators $\hat{c}_{k,\sigma}$ are usually defined 
using Bloch states
[i.e., the one-electron eigenstates
of the Hamiltonian~(\ref{Hamiltonian}) with 
periodic boundary conditions $\hat{c}_{l+L,\sigma}
=\hat{c}_{l,\sigma}$ in the non-interacting limit ($U=0$)]:
$\hat{c}_{k,\sigma} = L^{-1/2}
\sum_{l} e^{-ikl} \hat{c}_{l,\sigma}$
with momentum $k=2\pi z/L$ for integers $-L/2<z\leq L/2$. 
Since DMRG calculations can be performed for much larger systems 
using open boundary conditions, it is desirable
to extend the definition of the spectral function $A(k,\omega)$
to that case.
Therefore, we use the eigenstates of the particle-in-a-box 
problem [i.e., the one-electron eigenstates of
the Hamiltonian~(\ref{Hamiltonian}) 
on an open chain for $U=0$] to
define the operators  
\begin{equation}
\hat{c}_{k,\sigma} = \sqrt{\frac{2}{L+1}} \sum_{l} \sin(kl)
\hat{c}_{l,\sigma}
\label{quasimom}
\end{equation}
with (quasi-)momentum $k=\pi z/(L+1)$ for integers $ 1 \leq z \leq L$.
Both definitions of $\hat{c}_{k,\sigma}$
should be equivalent in the thermodynamic limit
$L\rightarrow \infty$.
Tests for finite systems up to $L=32$ sites show that
both approaches are consistent 
except in the asymptotic Luttinger
liquid regime [i.e.,
at low energy ($|\omega| \sim 1/L$)
close to the Fermi vector $k_F = \pi n/2$ 
($|k-k_F| \sim 1/L$)].
Therefore, open chains and the definition~(\ref{quasimom})
can be used to investigate the spectral function $A(k,\omega)$.
In this paper we present only results obtained using this approach.

\begin{figure}
\includegraphics[width=6.6cm]{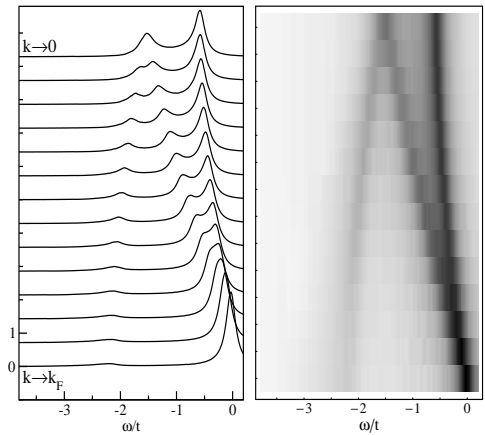}
\caption{ \label{fig1}
Line shapes (left) and gray-scale plot (right) of the
spectral function $A(k,\omega)$ for $0<k<k_F$
calculated with a broadening $\eta = 0.1t$ using DDMRG.
}
\end{figure}

\begin{figure}
\includegraphics[width=6.6cm]{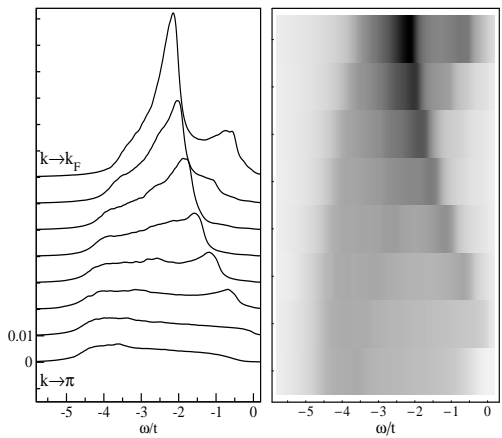}
\caption{\label{fig2}
Line shapes (left) and gray-scale plot (right) of the
spectral function $A(k,\omega)$ for $k_F < k < \pi$
calculated with a broadening $\eta = 0.1t$ using DDMRG.
}
\end{figure}

Figures~\ref{fig1} and~\ref{fig2} show the spectral function
calculated with DDMRG in a chain with $L=90$ sites.
Since the spectrum is symmetric, $A(-k,\omega) = A(k,\omega)$,
we show results for $k \geq 0$ only.
Three dispersing features are clearly visible in the spectrum 
for $|k| < k_F$ in Fig~\ref{fig1}.
At small binding energy $-\omega$ there are intense peaks
with a narrow dispersion (from $\omega \approx 0$
at $k=\pm k_F$ to $\omega \approx -0.5t$ at $k=0$).
This feature corresponds to the 
spinon branches in the Luttinger liquid regime.
Note that both spinon branches (for $k < 0$ and $k > 0$)
join at $k=0$ and thus form just one spinon band.
At energies $\omega$ lower than the spinon band
there is a second spectral feature made of peaks with less spectral weight
and a wider dispersion (from $\omega\approx 0$ at $k=\pm k_F$
to $\omega \approx -1.5t$ at $k=0$).
It merges with the spinon band for
$k \rightarrow \pm k_F$ because of the finite broadening.
This feature corresponds to the two holon branches 
of the Luttinger liquid theory. 
The third spectral feature is made of weaker peaks and has
an (apparently) inverted 
dispersion (starting at $\omega \approx -1.5t$ for $k=0$
and reaching $\omega \approx -2.2t$ at $k=\pm k_F$).
These so-called shadow bands~\cite{Penc} are actually the
continuation of the holon bands.
Thus the second and third features correspond
to two holon/shadow bands crossing at $k=0$.
While the spectral weight of the structure associated
with the spinon and holon bands
remains relatively constant for all $|k| < k_F$,
the shadow bands rapidly loose intensity with increasing
$k$.

For $|k| > k_F$ the spectral weight is much
lower than for $|k| < k_F$ (see Fig.~\ref{fig2}).
Nevertheless, one can observe four dispersive structures
in the spectral function. 
First, the shadow band continues from $k=\pm k_F$ to
$\pm 3k_F$, but its energy increases with 
$|k|$ and approaches zero for $|k|=3k_F$. 
Weaker peaks are also visible at higher energy $\omega$
than the shadow band for $k_F < |k| < 2k_F$.
The corresponding binding energy $-\omega$
increases from about zero at
$k = \pm k_F$ to about $1.7t$ at $k = \pm 2k_F$, where
this second feature meets the shadow band and 
apparently disappears.
The third dispersing feature corresponds to very weak peaks
(not visible on the scale of
Fig.~\ref{fig2})
with energies from $\omega \approx -3.7t$ at $k=\pi$
to $\omega \approx = -2.2t$ at $|k| \approx 2k_F$.
Note that, despite its weakness,
this feature corresponds to the spectrum maximum for $k\approx\pi$.
The last feature is a sharp drop of the spectral weight at low energy. 
It goes from $\omega \approx -3.25t$ at
$|k| \approx k_F$ to $\omega \approx -4.6t$ at $k=\pi$.
We interpret this drop as the lower edge of the 
photoemission spectrum.
The little spectral weight found at lower energy $\omega$
is due to the finite broadening $\eta$
used in our DDMRG calculations. 
Note that these third and fourth spectral features
are not visible for small $|k|$ because
they are too close to the broadened and comparably much
stronger peaks belonging to other structures.

\begin{figure}
\includegraphics[width=8cm]{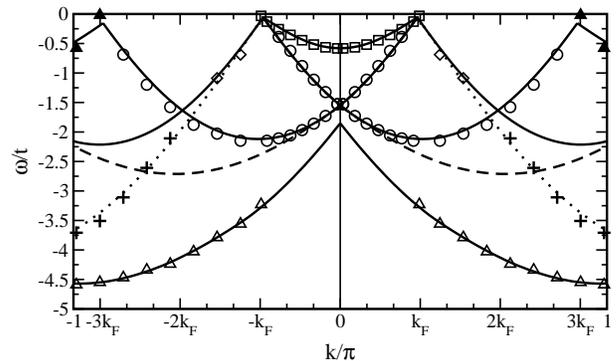}
\caption{\label{fig3}
Dispersion $\omega(k)$ of the structures observed in the DDMRG
spectral function: spinon band (square), holon-shadow bands
(circle), secondary holon bands (diamond), lowest ``4$k_F$''-singlet
excitations (plus), and lower 
(open triangle) and upper (solid triangle) spectrum edges. 
Lines show dispersions $\epsilon(k)$ obtained from the Bethe
Ansatz solution.
}
\end{figure}

Figure~\ref{fig3} shows the dispersion $\omega(k)$ of the various
features found in the DDMRG spectrum for the 90-site chain.
One clearly sees that
the shadow bands are just the continuation of
the holon bands. 
The dispersions $\omega(k)$ 
should naturally correspond to specific excitation bands 
$\epsilon(k)$ of the Hubbard model.
To identify these excitations 
we have calculated the excitation energies
$\epsilon(k)$ for the removal of an electron
in the Hubbard model on a 90-site chain 
using the Bethe Ansatz solution~\cite{Schulz}.
In Fig.~\ref{fig3} we show those excitation bands $\epsilon(k)$ which
correspond to the dispersing features found
in the DDMRG spectral function.
The excellent  quantitative agreement between the 
Bethe Ansatz results calculated 
for periodic boundary conditions and our numerical data
confirms
that the open chains used in our DDMRG calculations
do not affect the spectral properties significantly.

Due to the separation of spin and charge dynamics, electron-removal
excitations with momentum $k$ are made of independent
spin and charge excitations with momenta
$k_s$ and $k_c = k-k_s$, respectively.
The spinon band between $-k_F$ and $k_F$ is related to
excitations with the lowest possible binding energy 
for $k_c = 0$ and $|k_s| \leq k_F$.
This defines the spinon dispersion $\epsilon_s(k_s)$,
which has a width of about $0.5t$ and  
gives the spectral onset for $|k| < k_F$.  
The holon/shadow bands going from $-k_F$ to $3k_F$
and from $-3k_F$ to $k_F$ correspond to  
excitations with the lowest possible binding energy
for $|k_s| =  k_F$, $0 \leq |k_c| \leq 4k_F$,
and $k_s k_c < 0$.    
This defines the holon dispersion $\epsilon_c(k_c)$
with a width of about $2t$. 
It gives the spectral onset for $2k_F \leq |k| \leq 3k_F$.
The peaks found at low binding energy for $k_F \leq |k| \leq 2k_F$ 
correspond to secondary holon bands made of similar 
excitations as the holon-shadow bands but 
with parallel spin and charge momenta ($k_s k_c > 0$).
They give the spectral onset for $k_F \leq |k| \leq 2k_F$.
For $3k_F < |k| < \pi$ this onset corresponds to
a secondary spinon band with $k_c = \pm 4k_F$ and 
$|k_s| < k_F$.

In Fig.~\ref{fig3}, a dashed line shows the
dispersion of the lowest possible excitations made of one spinon
and one holon [i.e, the minimum of $\epsilon_c(k_c)+
\epsilon_s(k_s)$ for a given $k=k_c+k_s$].
This lower edge of the spinon-holon continuum is not related
to any feature in the DDMRG spectral function and one finds
spectral weight at lower energy $\omega$.
Therefore, the Hubbard model spectral function can not be
explained with spinon-holon excitations only.
Actually, the lower edge of the spectrum follows
the dispersion of the lowest states made of one spinon and a single
charge excitation called ``$4k_F$''-singlet excitation in
Ref.~\cite{Schulz}. 
Finally, the very weak peaks found for $-2t > \omega > -4t$ and $|k| \agt 2k_F$
seem to be related to the lowest possible ``4$k_F$''-singlet charge excitations
with $k_s=\pm k_F$ and  $k_c k_s > 0$.

In Ref.~\cite{Sing} it was shown that the dispersion 
of the TCNQ related peaks in the ARPES spectrum of
TTF-TCNQ could be mapped onto excitation bands
of a one-dimensional Hubbard model.  
Our DDMRG calculations show that the Hubbard model
also explains qualitatively the experimental
spectral weight distribution.
(A quantitative comparison is not possible because
of the strong background contribution in the
ARPES data.)
The ARPES spectrum features labeled (a), (b), and
(d) in Refs.~\cite{Claessen,Sing} perfectly match 
the (singular) features found in the Hubbard model
spectral function (the spinon, holon, and shadow bands,
respectively). 
This confirms that the ARPES spectrum of TTF-TCNQ
shows the signature of spin-charge separation 
over the scale of the conduction band width  
(of the order of $1eV$).
In addition, we note that the secondary
holon bands (for $k_F < |k| < 2 k_F$) correspond
to a poorly understood spectral feature [labeled (c)] 
which has been attributed to excitations
of the TTF band in Refs.~\cite{Claessen,Sing}.
Therefore, we think that this spectral feature
is not related to the TTF band but is 
naturally explained by the TCNQ secondary holon bands,
at least in the range $k_F < |k| < 2 k_F$.

In the Luttinger liquid theory
the spectral functions $A(k,\omega)$ have singularities
$|\omega-\epsilon(k)|^{-\alpha}$ for energies
$\epsilon(k) \propto |k \pm k_F|$ given by the spinon and holon 
linear dispersions~\cite{Voit,Meden}.
For a system which is invariant under spin rotation
the exponents $\alpha$ are related to the 
Luttinger liquid parameter $K_{\rho}$ through
$\alpha_s = (4-K_{\rho}-K^{-1}_{\rho})/4$ on the
spinon branch and 
$\alpha_c = (6-K_{\rho}-K^{-1}_{\rho})/8$ 
on the holon branch.  
The parameter $K_{\rho}$ can be calculated
in the one-dimensional Hubbard model~\cite{Schulz2}
and one finds $K_{\rho} \approx 0.68$ 
for $U=4.9t$ and $n=0.6$, which corresponds
to exponents $\alpha_s = 0.46$ and
$\alpha_c = 0.48$.

\begin{figure}
\includegraphics[width=6.5cm]{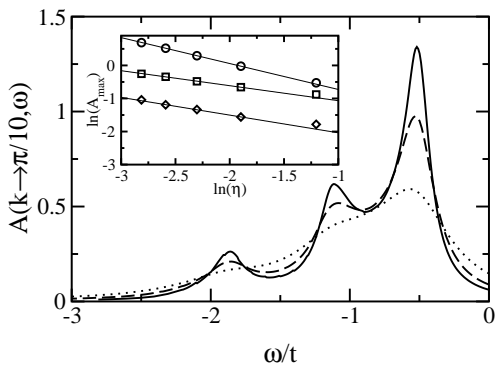}
\caption{\label{fig4}
Spectral functions $A(k \approx \pi/10=k_F/3,\omega)$ calculated with 
DDMRG for system sizes $L=30$ (dotted), 60 (dashed), and 90 (solid).
Inset: scaling of the peak maxima for $0.3 \leq \eta/t \leq 0.06$
($30 \leq L \leq 150$). Solid lines are fits.
}
\end{figure}

In view of the Luttinger liquid theory results
it is natural to ask whether the broadened
peaks found in our DDMRG calculations 
become singularities of the spectral
function in the thermodynamic limit.  
To answer this question and to estimate the
exponents $\alpha$ we have performed a
finite-size-scaling analysis~\cite{ddmrg}. 
The spectral function $A(k,\omega)$ is calculated
for several system sizes $L$ with a broadening
scaling as $\eta=9t/L$.
Some spectra for $k \approx \pi/10 = k_F/3$
are shown in Fig.~\ref{fig4}.
The scaling of the peak maxima $A_{\text{max}}$ with $\eta$
can then be analyzed (see the inset of Fig.~\ref{fig4}). 
If $A_{\text{max}}$ diverges as $\eta^{-\alpha}$
($0 < \alpha < 1$)
for $\eta \rightarrow 0$, the spectral function
has a singularity with exponent $\alpha$ in the
thermodynamic limit.
A Landau quasi-particle 
corresponds to a Dirac $\delta$-function
and thus to a peak diverging as 
$\eta^{-1}$. 

Using this scaling analysis
we have found that 
the spinon, holon and shadow band peaks
become singularities
in the thermodynamic limit.
We have not found any 
diverging peak with an exponent larger
than 0.86,
which confirms the absence of Landau
quasi-particles.
For $k = \pi/10$ we have found that the
spinon, holon, and shadow band
exponents are
$\alpha =$ 0.78, 0.44, and 0.56, respectively.
For $k=0$, we have obtained
$\alpha = 0.86$ for the spinon band
and $\alpha = 0.70$ for the holon/shadow band.
Therefore, the exponents $\alpha$ are 
momentum-dependent and for finite $|k\pm k_F|$ 
they are significantly different
from the Luttinger liquid predictions for $|k| \rightarrow k_F$.
A recent study~\cite{Carmelo}
has also shown that these exponents are strongly
$k$-dependent.
It is not possible to 
determine the exponents $\alpha$ in the
asymptotic Luttinger liquid regime
with DDMRG because 
the finite-size effects are not under control
in that limit.

In summary, we have used a novel approach to compute
the photoemission spectral function of the Hubbard model
on open chains using DDMRG and explained the
ARPES spectrum of the organic conductor TTF-TCNQ. 
Our method can easily be extended to other dynamical response functions and 
more complicated models.

We are grateful to R.~Claessen, F.H.L.~E\ss ler, and J.~Carmelo
for helpful discussions. H.B. acknowledges 
support by the Optodynamics Center of the
Philipps-Universit\"{a}t Marburg and
thanks the Institute for Strongly Correlated and Complex
Systems at Brookhaven National Laboratory
for its hospitality. 


\begin{thebibliography}{99}

\bibitem{Schoenhammer} K. Sch\"{o}nhammer, 
\textit{Luttinger liquids: the basic concepts},
to be published in \textit{Strong Interactions in 
Low Dimensions}, edited by D. Baeriswyl and L.~Degiorgi
(Kluwer, 2004); e-print cond-mat/0305035 (unpublished).

\bibitem{Voit} J. Voit, \prb \textbf{47}, 6740 (1993).

\bibitem{Meden} V. Meden and K. Sch\"{o}nhammer, \prb \textbf{46},
15753 (1992).

\bibitem{Claessen} R. Claessen \textit{et al.}, \prl
\textbf{88}, 096402 (2002).

\bibitem{Sing} M. Sing \textit{et al.}, \prb \textbf{68},
125111 (2003).

\bibitem{Hubbard} J.~Hubbard, Proc.~R.~Soc.~London A \textbf{276}, 
238 (1963).

\bibitem{Lieb} E.H.~Lieb and F.Y.~Wu, \prl \textbf{20},
1445 (1968).

\bibitem{Schulz} H.J. Schulz, e-print cond-mat/9302006 (unpublished).

\bibitem{Penc} K. Penc, K. Hallberg, F. Mila, and H. Shiba,
\prl \textbf{77}, 1390 (1996); \prb \textbf{55}, 15475 (1997).

\bibitem{Favand} J. Favand \textit{et al.}, \prb \textbf{55},
4859 (1997).

\bibitem{Bannister} R.N. Bannister and N.~d'Ambrumenil,
\prb \textbf{61}, 4651 (2000).

\bibitem{QMC} R. Preuss \textit{et al.}, \prl \textbf{73},
732 (1994); M.G.~Zacher, E.~Arrigoni, W.~Hanke, and 
J.R.~Schrieffer, \prb \textbf{57}, 6370 (1998).

\bibitem{Senechal} D. S\'{e}n\'{e}chal, D. Perez, 
and M.~Pioro-Ladri\`{e}re, \prl \textbf{84}, 522 (2000).

\bibitem{ddmrg} E.~Jeckelmann, \prb \textbf{66}, 045114 (2002);
E.~Jeckelmann, F.~Gebhard, and F.H.L. Essler, \prl \textbf{85}, 
3910 (2000).

\bibitem{Schulz2} H.J. Schulz, \prl \textbf{64}, 2831 (1990).

\bibitem{Carmelo} J.M.P. Carmelo \textit{et al.}, Europhys. Lett.
(2004), to be published; e-print cond-mat/0307602 (unpublished).

\end{thebibliography}
\end{document}